\documentclass{aa}
\usepackage{graphicx}
\usepackage{longtable}
\usepackage{times}

\begin{document}

\title{A multi-transition molecular line study of candidate massive young stellar objects associated with methanol masers
\thanks{Figure A.1 and Table A.1 are only available in electronic form via
http://www.edpsciences.org} 
}

\author{M. Szymczak \inst{1}
        \and A. Bartkiewicz \inst{1}
        \and A.M.S. Richards \inst{2} 
        }

\institute{Toru\'n Centre for Astronomy, Nicolaus Copernicus 
          University, Gagarina 11, 87-100 Toru\'n, Poland 
   \and   Jodrell Bank Observatory, University of Manchester, 
          Macclesfield, Cheshire SK11 9DL, UK
          }
\date{Received 13 February 2007 / Accepted 19 March 2007 }

\titlerunning{Molecular line study of high-mass protostars}

\authorrunning{M. Szymczak et al.}

\abstract {}{We characterize the molecular environment of candidate
massive young stellar objects (MYSOs) signposted by methanol masers.}
{Single pixel observations of 10 transitions of HCO$^+$, CO and CS
isotopomers were carried out, using the IRAM 30\,m telescope.  We
studied a sample of 28 targets for which the 6.7\,GHz maser emission
positions are known with a sub-arcsecond accuracy.}  {The systemic
velocity inferred from the optically thin lines agrees within
$\pm$3\,km\,s$^{-1}$ with the central velocity of the maser emission 
for most of the sources. About 64\% of the sources show
line wings in one or more transitions of CO, HCO$^+$ and CS species,
indicating the presence of molecular outflows.  Comparison of the
widths of line wings and methanol maser emission suggests that the
6.7\,GHz maser line traces the environment of MYSO of various
kinematic regimes. Therefore conditions conducive for the methanol
maser can exist in the inner parts of molecular clouds or circumstellar 
discs as well as in the outer parts associated with molecular
outflows. Calculations of the physical conditions based on
the CO and HCO$^+$ lines and the CS line intensity ratios refine the
input parameters for maser models.  Specifically, a gas number density 
of $<10^7$\,cm$^{-3}$ is sufficient for strong maser emission and 
a high methanol fractional abundance ($>5\times10^{-7}$) is required.}
{} 

\maketitle

\section{Introduction}
There is compelling evidence that methanol masers are a signature of
recent or ongoing high-mass star formation (Menten\,\cite{menten91}).
However, it is not yet fully understood when they appear in an
evolutionary sequence and what they actually trace.  The evaporation
of grain mantles is postulated as the main process enhancing the 
fractional abundance of methanol molecules in the gas phase up to 
$10^{-6}$ (Dartois et al.\,\cite{dartois99}). This implies that
methanol masers can emerge after the formation of an embedded heating 
source.  Methanol maser sources rarely show strong ($>$100\,mJy) 
free-free emission at centimetre wavelengths implying that they
precede the development of detectable ultra-compact HII (UCHII) region 
(Walsh et al.\,\cite{walsh98}; Codella \&
Moscadelli\,\cite{codellamosca00}). 
The estimated lifetime of methanol masers is a few $\times10^4$\,yr 
(van der Walt\,\cite{vanderwalt05}) which is similar to the typical 
dynamical timescales of molecular outflows. High spatial resolution 
observations revealed a variety of maser site sizes from 40$-$1200\,AU 
(e.g. Norris et al.\,\cite{norris98}; Walsh et
al.\,\cite{walsh98}; Minier et al.\,\cite{minier00}). The maser
emission arises either from circumstellar discs or behind shocks
tracing outflows from massive young stellar objects (MYSOs). No object
was found that unequivocally confirms one of these scenarios.

The non-linear nature of maser amplification means that it is
difficult to relate the maser line intensity directly to the physical
parameters of the active region. Theoretical models predict the
formation of methanol maser lines under a rather wide range of gas
and dust temperatures (30$-$200\,K and 100$-$300\,K, respectively) 
and hydrogen number densities ($10^5-10^8$\,cm$^{-3}$)
(Cragg et al.\,\cite{cragg02}).  Thus, it appears that a better
understanding of the environments in which the masers arise is
required in order to realise their full potential as probes of
the formation of high-mass stars.

In this paper we report our attempts to constrain the range of
environments probed by methanol masers using observations of thermal
emission from  other molecular species and lines. Specifically, 
the ratios of the intensities of different transitions of CS and 
C$^{34}$S molecules are used to obtain the temperature and density 
of the gas.  The optically thin and thick lines of CO and HCO$^+$ are
used to constrain the column density. These techniques 
were successfully used to characterize other samples
of MYSOs (e.g. Plume et
al.\,\cite{plume97}; Beuther et al.\,\cite{beuther02a}; Purcell et
al.\,\cite{purcell06}).
Additionally, the molecular line profiles yield information on the
kinematics of various parts of the molecular clouds surrounding the
high-mass protostars (e.g. Fuller et al.\,\cite{fuller05}; Purcell et
al.\,\cite{purcell06}).
 
A homogeneous and unbiased sample of MYSOs is necessary in order 
to address these issues properly.  Our recent 6.7-GHz unbiased survey 
for methanol masers in selected regions of the Galactic plane 
(Szymczak et al.\,\cite{szymczak02})
provides such a complete, sensitivity limited sample of candidate MYSOs.
Objects identified in the survey probably represent a class of
MYSOs in an early evolutionary phase. Some groups and individual
sources in this class, selected using various diagnostics of high-mass
star formation, have been studied in thermal molecular lines (Brand et
al.\,\cite{brand01}; Beuther et al.\,\cite{beuther02a}; Fuller et
al.\,\cite{fuller05}), but this is the first published study of a
homogeneous sample based solely on the presence of detectable methanol
masers.

\section{The sample}
The 28 sources observed in this study (Table 1) were chosen from a
sample of 100 methanol maser sources found in the Torun 32\,m
telescope blind survey for the 6.7\,GHz methanol line in the Galactic
plane area $20^\circ\leq l\leq 40^\circ$ and $|b|\leq 0.^\circ 52$
(Szymczak et al.\,\cite{szymczak02}).  This flux-limited
($3\sigma \simeq$1.6\,Jy) subsample includes 25 out of 26 sources 
which were undetected prior to the Torun survey.
Therefore, our subsample specifically excludes previously 
known sources associated with OH maser emission 
(Caswell et al.\,\cite{caswell95}) or with  IRAS-selected
bright UCHII candidates (Schutte et al.\,\cite{schutte93}; 
van der Walt et al.\,\cite{vanderwalt95};
Walsh et al.\,\cite{walsh97}).  Assuming that CH$_3$OH 
masing precedes the appearance of OH masers and detectable UCHII
regions, the objects studied here represent sites of high-mass 
star formation at a very early stage. The average peak maser 
flux of the 28 targets is 17.3\,Jy, a factor of 2.6 lower than 
that of the other 72 objects in the original sample, suggesting 
that distant or intrinsically faint objects may be 
over-represented in our subsample. The subsample studied here 
is most certainly not complete.

%====================================================================================
\begin{table*}[t]
\caption{List of targets.}
\begin{tabular}{l l l l l l l}
\hline
Name & $\alpha$(J2000)  & $\delta$(J2000) & $\sigma_{\alpha}$
& $\sigma_{\delta}$ & $V_{\rm p}$ & $S_{\rm p}$ \\
     &               &            &    (arcsec) & (arcsec) &
(km\,s$^{-1}$) & (Jy)  \\
\hline
21.407$-$0.254 & 18 31 06.3403 & $-$10 21 37.305 & 0.28 & 0.80 & +89.0  & 2.0 \\%t1 5.2
22.335$-$0.155 & 18 32 29.4109 & $-$09 29 29.435 & 0.27 & 1.10 & +35.7  & 2.8 \\%new 28.1
22.357$+$0.066$^1$ & 18 31 44.144  & $-$09 22 12.45  &   &  &  +80. &  \\%known 38.7
23.707$-$0.198$^3$ & 18 35 12.3625 & $-$08 17 39.409 & 0.06 & 0.40 & +79.0  & 3.2 \\%t1 7.8
23.966$-$0.109$^3$ & 18 35 22.2167 & $-$08 01 22.395 & 0.35 & 1.60 & +71.0  & 4.3 \\%t1 13.1
24.147$-$0.009$^3$ & 18 35 20.9501 & $-$07 48 57.470 & 0.03 & 0.19 & +17.9  & 6.4 \\%t1 25.8  
24.541$+$0.312$^3$ & 18 34 55.7212 & $-$07 19 06.630 & 0.90 & 0.90 & +105.5 &  4.4  \\%t1 12.2
24.635$-$0.323 & 18 37 22.7932 & $-$07 31 37.950 & 0.50 & 1.20 & +35.5  & 1.0 \\%t1 2.3
25.410$+$0.105$^2$ & 18 37 16.9 & $-$06 38 30.4  &  &  & +97.    & \\%known  3.7
26.598$-$0.024 & 18 39 55.9268 & $-$05 38 44.490 & 0.03 & 0.18 & +23.0  & 2.0 \\%t1 18.4
27.221$+$0.136 & 18 40 30.5446 & $-$05 01 05.450 & 0.03 & 0.18 & +119.0 & 3.0 \\%new 9.9
28.817$+$0.365 & 18 42 37.3470 & $-$03 29 41.100 & 0.02 & 0.18 & +91.0  & 1.0 \\%new  7.2
30.316$+$0.069 & 18 46 25.0411 & $-$02 17 45.160 & 0.03 & 0.16 & +35.5  & 1.3 \\%t1  8.5
30.398$-$0.297 & 18 47 52.2623 & $-$02 23 23.660 & 0.02 & 0.14 & +98.2  & 1.5 \\%new  5.1
31.056$+$0.361 & 18 46 43.8558 & $-$01 30 15.690 & 0.05 & 0.28 & +81.0  & 1.0 \\%t1  3.7
31.156$+$0.045 & 18 48 02.3471 & $-$01 33 35.095 & 0.10 & 0.90 & +41.0  &  0.8 \\%t1  1.9
31.585$+$0.080 & 18 48 41.8975 & $-$01 09 43.085 & 0.50 & 0.70 & +95.8  & 0.8 \\%t1  3.1
32.966$+$0.041$^2$ & 18 51 24.5 & $+$00 04 33.7  &  &  &  +92.  & \\%known   30.7
33.648$-$0.224$^3$ & 18 53 32.5508 & $+$00 32 06.525 & 0.50 & 1.0  & +62.6   & 20.0 \\%t1  108.7
33.980$-$0.019 & 18 53 25.0184 & $+$00 55 27.260 & 0.05 & 0.50 & +59.0   & 1.0 \\%t1  3.8
34.753$-$0.092 & 18 55 05.2410 & $+$01 34 44.315 & 0.08 & 0.50 & +53.0   & 1.6 \\%new  9.7
35.791$-$0.175$^3$ & 18 57 16.9108 & $+$02 27 52.900 & 0.04 & 0.17 & +60.8   &  5.6 \\%new  24.5
36.115$+$0.552$^3$ & 18 55 16.8144 & $+$03 05 03.720 & 0.02 & 0.23 & +74.2   & 7.2 \\%t1  43.3
36.704$+$0.096 & 18 57 59.1149 & $+$03 24 01.395 & 0.08 & 0.17 & +53.0   &  1.9 \\%new  8.6
37.030$-$0.039 & 18 59 03.6435 & $+$03 37 45.140 & 0.14 & 0.50 & +79.0   &  1.2 \\%new 7.3
37.479$-$0.105 & 19 00 07.1457 & $+$03 59 53.245 & 0.07 & 0.36 & +62.8   & 1.8 \\%t1  12.5
37.600$+$0.426 & 18 58 26.8225 & $+$04 20 51.770 & 0.03 & 0.70 & +91.2   & 2.0 \\%new 24.3
39.100$+$0.491$^3$ & 19 00 58.0394 & $+$05 42 43.860 & 0.34 & 0.17 & +15.2   & 2.8 \\%new  15.3
\hline 
\end{tabular}
\label{pos}

$^1$ Position is from Walsh et al. 1998, $^2$ Position is from Beuther et al. 2002a,
$^3$ This source was reported in Niezurawska et al. 2005
\end{table*}
%====================================================================================

\subsection{Astrometric positions}
The coordinates and position uncertainties of the brightest
6.7-GHz maser component in each source are presented in Table 1.
The LSR velocity of this component ($V_{\rm p}$) and its peak flux
density ($S_{\rm p}$) are given for each target. 
The positions and flux densities of all but three objects were 
measured with the Mark II $-$ Cambridge baseline of MERLIN in 
two sessions between 2002 May and 2003 May. For the three objects
not measured the peak velocities were taken from Szymczak et al.
(\cite{szymczak02}).

The observational setup and data reduction were described in Niezurawska
et al. (\cite{niezurawska05}). A primarily goal of those astrometric
measurements was to determine the positions with sub-arcsecond
accuracy for follow-up VLBI observations.  Measurement errors mainly
depended upon the ratio of the beam size to the signal to the noise
ratio (Thompson et al.\,\cite{thompson91}).  
If the emission was complex we took the dispersion of neighbouring 
maxima as the position uncertainty. The N$-$S elongation of the 
synthesized beam  close to declination 0\degr ~produces a split peak, 
in which case the position uncertainty in that direction was taken 
as half the separation of the maxima.
Consequently, for sources with a single clear peak, the position
errors in right ascension were as small as 0\farcs02 but increased up
to 0\farcs90 for sources with complex emission.  The respective errors
in the declination were 0\farcs14 and 1\farcs6. Comparison with our
unpublished VLBI observations reveals position differences between
MERLIN and VLBI measurements no worse than a few tens of
milli-arcseconds.  This implies that the values listed in Table 1 are
maximal position errors for most of the targets. The flux densities
listed in Table 1 are a factor of 2$-$3 lower than those measured with
the single dish (Szymczak et al.\,\cite{szymczak02}) and should be
considered as lower limits. 
The exact flux scale and gain-elevation effects for low-declination 
sources are not yet fully investigated at 6 GHz but comparison of 
calibration sources in common with other experiments shows that 
the uncertainties are 10 $- \le$50\%. This suggests that about 
half the methanol flux arises on scales larger than the beam size 
of 50$-$100 mas.

\subsection{Distances}
The distances were determined using the Galactic rotation curve
of Brand \& Blitz (\cite{brand93}) and the central velocity
of each 6.7\,GHz methanol maser profile as measured
by Szymczak et al. (\cite{szymczak02}). Selection of this
velocity as a reliable estimator of the systemic velocity is proven  
in Sect. 6.1. The sources are all in the first quadrant so that there
is an ambiguity between the near and far kinematic distances. 
In most cases we are unable to resolve this ambiguity because there 
are no independent distance measurements in the literature for
our sample. Based on the arguments discussed in 
Walsh et al. (\cite{walsh97}), we adopted the near kinematic distances
(Table 4).

\section{Observations and data reduction}
Observations were carried out between 2004 September 28 and
October 2 with the IRAM 30\,m telescope. Ten transitions of HCO$^+$,
CO and CS isotopomers were observed. Two or three SIS receivers
tuned to single sideband mode were used simultaneously, in combination 
with the VESPA autocorrelator as well as with 100\,kHz and 1\,MHz filter
banks. Table 2 lists the rest line frequencies, half power beam widths
(HPBWs), velocity resolutions and typical system temperatures for each 
transition.

The data were taken using the position switching mode. The off
positions were typically 30\arcmin\, away from the targets. 
In the few cases, especially for the C$^{13}$O J=2$-$1 line, where emission 
was seen at the reference position, the offsets were increased 
up to 45\arcmin\, in the direction away from the Galactic plane. The observations
were centered on the target positions listed in Table 1.
Integration times on-source in total power mode were 12$-$18 min
per frequency setting, resulting in rms noise levels ranging from
$\approx$0.05\,K at 87\,GHz to $\approx$0.90\,K at 245\,GHz for 
a spectral resolution of 0.10$-$0.16\,km\,s$^{-1}$.
Pointing was checked regularly on nearby continuum sources and
was usually found to be within 2\arcsec ~and always within 3\arcsec.
The spectra were scaled to the main beam brightness temperature 
($T_{\rm mb}$) using the efficiencies supplied by the
observatory{\footnote{http://www.iram.es/IRAMES/telescope/telescopeSummary/telesco-
 pe\_summary.html}}. 
Comparison of our data with those taken by
Brand et al. (\cite{brand01}) with the same telescope and spectral
resolutions for a source in common, 36.115+0.552, implies consistent 
intensity scales within less than 30\%.  

The data reduction were performed using the CLASS software package.
Low order polynomials were applied to remove baselines from
the calibrated spectra. The line parameters were determined from
Gaussian fits and are listed in Table A.1 where the following
information is given: the rms (1$\sigma$) noise level, 
the extreme velocities $V_{\rm s}$, $V_{\rm e}$ where the intensity 
drops below the 2$\sigma$ level, the peak temperature $T_{\rm mb}$,
the velocity of the peak $V_{\rm p}$,
the line width at half maximum $\Delta V$ and the integrated line
intensity $\int T_{\rm mb}{\rm d}v$. Velocities are in the LSR frame.
In several cases where the profiles were non-Gaussian, these values 
were read off from the spectra.
In some cases the spectra were smoothed to improve the signal to noise
ratio. In this paper, only the autocorrelation spectra are analyzed. 

%====================================================================================
\begin{table}
\caption {Observing parameters}
\begin{tabular}{l l c c c l}
\hline
 Transition   &\hskip1mm Frequency & Ref.   &\hskip-1mm HPBW      &\hskip-1mm Res. &\hskip0.5mm T$_{\rm sys}$ \\
              & \hskip3mm (GHz)    &        &\hskip-1mm (\arcsec) &\hskip-1mm  (km\,s$^{-1}$) &\hskip1mm  (K)      \\
\hline
HCO$^{+}$(1$-$0) &\hskip1.5mm  89.188518 &  2  &\hskip-1mm 27       &\hskip-1mm 0.13  &\hskip1.5mm   200  \\
H$^{13}$CO$^{+}$(1$-$0) &\hskip1.5mm  86.754330 & 1 &\hskip-1mm 27 &\hskip-1mm  0.13  &\hskip1.5mm   200  \\
$^{13}$CO(2$-$1) & 220.398686 &   2 &\hskip-1mm 11      &\hskip-1mm  0.11  &  1200  \\
C$^{18}$O(2$-$1) & 219.560328 &   2 &\hskip-1mm 11      &\hskip-1mm  0.11  &  1340  \\
CS(2$-$1)        &\hskip1.5mm 97.980953  &   1 &\hskip-1mm 25      &\hskip-1mm  0.12  &\hskip1.5mm   260  \\
CS(3$-$2)        & 146.969049 &   2 &\hskip-1mm 17      &\hskip-1mm  0.16  &\hskip1.5mm   690  \\
CS(5$-$4)        & 244.935560 &   1 &\hskip-1mm 10      &\hskip-1mm  0.10  &  1830  \\
C$^{34}$S(2$-$1) &\hskip1.5mm 96.412961  &  1 &\hskip-1mm 25      &\hskip-1mm  0.12  &\hskip1.5mm   290  \\
C$^{34}$S(3$-$2) & 144.617147 &   2 &\hskip-1mm 17      &\hskip-1mm  0.16  &\hskip1.5mm   560  \\
C$^{34}$S(5$-$4) & 241.016113 &   1 &\hskip-1mm 10      &\hskip-1mm  0.10  &  2100  \\
\hline
\end{tabular}
\label{observing}

The references for the line frequencies are 1 - Lovas (2003), 
2 - Brand et al. (2001)
\end{table}

%====================================================================================

\section{Results}
The basic parameters of molecular transitions derived from Gaussian
analysis are assembled in Table A.1, while the spectra are shown in 
Fig. A.1. 

The sensitivity achieved and detection rate for each transition are
summarized in Fig. 1. The histogram counts as detected only those
lines with $T_{\rm mb}>3\sigma$. $^{13}$CO(2$-$1), C$^{18}$O(2$-$1),
CS(2$-$1) and CS(3$-$2) lines were detected in all sources. 
HCO$^+$(1$-$0) and H$^{13}$CO$^+$(1$-$0) lines were detected in all
but one source. The detection rates in C$^{34}$S(2$-$1) and C$^{34}$S(3$-$2)
transitions were about half of those in CS(2$-$1) and CS(3$-$2) lines.
Because the sensitivities achieved for these four lines were comparable,
these detection rate differences reflect a real drop in the number of
sources exhibiting emission at the same level in the C$^{34}$S(2$-$1)
and C$^{34}$S(3$-$2) lines.
In contrast, the lower detection rates in the CS(5$-$4) and C$^{34}$S(5$-$4)
transitions appear to reflect the drop in sensitivity.

%===================================================================================
%===================================================================================
\begin{figure}[t]
    \resizebox{\hsize}{!}{\includegraphics[angle=0]{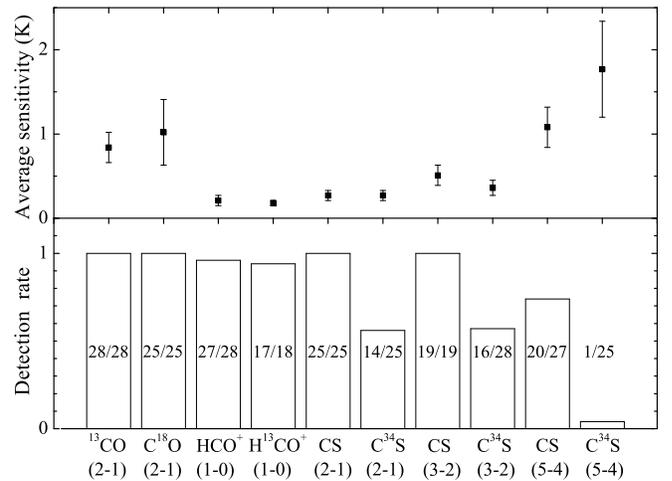}}
   \caption{The average sensitivity achieved for each transition
     (top) and the detection rate (bottom). The ratio of the number of
     detected to observed objects is shown in each of the bars.} 
  \label{figure1}
\end{figure}
%====================================================================================
%====================================================================================

\subsection{Systemic velocities}
Five of the observed lines (C$^{18}$O(2$-$1), H$^{13}$CO$^+$(1$-$0)
and the J=2$-$1, 3$-$2 and 5$-$4 transitions of C$^{34}$S) are
expected to be optically thin (Plume et al.\,\cite{plume97}; Brand et
al.\,\cite{brand01}; Purcell et al.\,\cite{purcell06}). These lines can be
used to determine source systemic velocities.  In order to test
whether these species trace the same or similar kinematic regimes we
compare their line parameters.  The C$^{34}$S(5$-$4) line is excluded
from the following analysis due to very low number of detections.

The average intensities of the H$^{13}$CO$^+$(1$-$0) and C$^{34}$S lines
are very similar and are a factor of 8 weaker than the average
intensity of the C$^{18}$O(2$-$1) line. This implies that
the parameters of the latter line, especially $V_{\rm p}$, are
determined most accurately.

We note that some line rest frequencies adopted from Brand et
al.\,(\cite{brand01}) differ slightly from those recommended by
Lovas (\cite{lovas03}). In the extreme case of C$^{34}$S(3$-$2) 
this results in the velocity difference of 0.07\,km\,s$^{-1}$. 
Moreover, the uncertainties in the line rest frequencies introduce 
a maximum uncertainty of $\pm$0.17\,km\,s$^{-1}$ to the derived 
line velocity for the H$^{13}$CO$^+$(1$-$0). We assume that the
above uncertainties affect the velocity estimates by up to
0.24\,km\,s$^{-1}$. Comparison of the velocities of the four
optically thin lines in our sample reveals no significant
average differences higher than 0.30\,km\,s$^{-1}$. This suggests
the same kinematic behaviour of these low-density gas tracers. 

At 100\,K the thermal linewidths of C$^{18}$O(2$-$1) and
H$^{13}$CO$^+$(1$-$0) are 0.24\,km\,s$^{-1}$ whereas those of
C$^{34}$S(2$-$1) and C$^{34}$S(3$-$2) are 0.20\,km\,s$^{-1}$.  
The observed linewidths are much broader, suggesting that turbulence or
bulk gas motions play a significant role in the line broadening.
The mean linewidth ratios of the optically thin lines are
5$-$10\% higher than unity. This bias is relatively small and suggests
that the lines trace the same molecular gas in the beam. 
The systemic velocities are listed in Table 4. They are primarily 
the C$^{34}$S(2$-$1) and C$^{34}$S(3$-$2) line peak velocities.
If emission in these lines is absent or weak the other optically
thin lines are used.  In two sources, 37.030$-$0.039 and 37.600+0.426, 
the systemic velocities are derived from CS(5$-$4) and HCO$^+$(1$-$0)
profiles, respectively.  We conclude that in most cases the observed 
optically thin lines are well fitted by single Gaussian profiles 
(deviations are discussed in Sect. 4.2.2) and their peak velocities 
are within $\pm$0.4\,km\,s$^{-1}$ of each other for
almost all sources in the sample.  Therefore, these lines provide
reliable estimates of systemic velocity of sufficient accuracy to allow
comparison with the methanol maser velocities listed in
Szymczak et al. (\cite{szymczak02}).

\subsection{Shape of profiles}
We analyse the molecular line profiles in order to search for specific
signatures of ordered motions such as infall, outflow or rotation.
Inward motion can be signposted by blue asymmetric profiles (Myers et
al.\,\cite{myers96}; Fuller et al.\,\cite{fuller05}) if the 
molecular spectral lines trace sufficiently dense gas. Signatures of
outflow or rotation are generally manifested in the line wings.

\subsubsection{Asymmetry}
We analysed line asymmetry quantitatively using the asymmetry
parameter (Mardones et al.\,\cite{mardones97}),
$\delta$v=(v$_{thick}$ $-$ v$_{thin}$)/$\Delta$V$_{thin}$, where
v$_{thick}$ and v$_{thin}$ are the peak velocities of optically
thick and optically thin lines, respectively and $\Delta$V$_{thin}$
is the line width at half maximum of the optically thin line.  
We used C$^{34}$S(2$-$1) as the optically thin line and the best
available measure of the systemic velocity of MYSOs.  Figure 2
shows histograms of the distribution of $\delta$v for the optically
thick lines $^{13}$CO(2$-$1), HCO$^+$(1$-$0), CS(2$-$1), CS(3$-$2) and
CS(5$-$4). There are approximately equal numbers of blue and red
asymmetric profiles in our sample.  Specifically, we note that there
is no evidence for an excess of blue-shifted emission in the
optically thick lines. Such an excess is postulated as the
signature of inward motion of the gas (Myers et al.\,\cite{myers96}). 
We suggest that motions other than infall, i.e. turbulence, rotation 
and outflow, are the dominant factor producing asymmetric profiles 
for most of the sources in our sample. It is possible that infall 
signatures could be masked by the relatively low resolution 
(typically $\ge 0.2$\,pc, i.e. at a distance of 5\,kpc and spatial
resolution of 10\arcsec) of our observations, since even the near 
kinematic distances are $>$3\,kpc for $\sim$80\% of the sources 
(the average $D_{\rm near}$ is $5.2\pm2.5$\,kpc for
the whole sample).

We therefore examined separately the 5 closest ($D_{\rm
near}<2.8$\,kpc) objects with well-determined asymmetry
parameters. Two of these, 26.598$-$0.024 and 30.316$+$0.069,
consistently show negative values of $\delta$v, i.e. blue asymmetry,
in the $^{13}$CO(2$-$1), HCO$^+$(1$-$0), CS(2$-$1) and CS(3$-$2) line profiles
(Fig. A.1). The corresponding values of $\delta$v are $-$0.38, $-$0.82,
$-$0.15 and $-$0.18 for source 26.598$-$0.024 and $-$0.43, $-$0.60,
$-$0.33 and $-$0.37 for source 30.316$+$0.069.  Their asymmetry
parameters are smaller for the optically thin lines
(i.e. C$^{18}$O(2$-$1) and H$^{13}$CO$^+$(1$-$0)), in the range from
$-$0.20 to 0.0. Such a dependence of the amount of blue asymmetry on the
optical depth of the transition is typical in molecular cores
experiencing infall (Narayanan et al.\,\cite{narayanan98}).   
We suggest that these two sources are the clearest
infall candidates although source complexity or a combination of outflow
and rotation could contribute to asymmetries in optically thick
profiles.

%===================================================================================
%===================================================================================
\begin{figure}[t]
    \resizebox{\hsize}{!}{\includegraphics[angle=0]{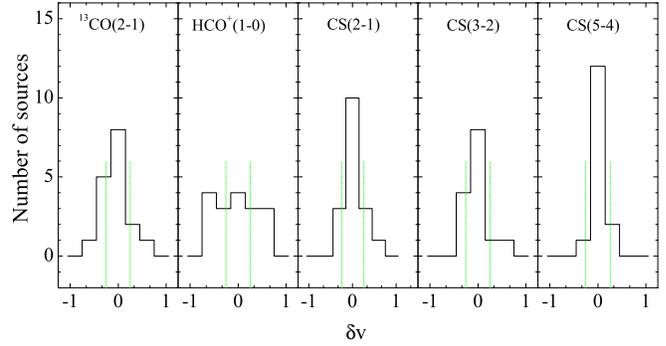}}
   \caption{Histograms of the distribution of the asymmetry parameter $\delta$v
     for the five transitions. The range of $|\delta$v$|< 0.25$ marked
     by the dotted lines corresponds to the spectra with no asymmetry.} 
  \label{figure2}
\end{figure}
%====================================================================================
%====================================================================================

\subsubsection{Wings}
Wing emission is identified by the presence of residuals after
Gaussian fitting and by comparing the same transitions of optically
thick and thin isopotomers.  A single Gaussian function provides a
good fit to most of the optically thin lines analyzed in Section 4.1, but
in a few cases the residuals are at a level $\ge3\sigma$, assumed to
be wing emission.  We cannot exclude the possibility that they are
weak separate component(s), given the limitations of our signal to
noise ratio and spectral resolution, but we note that the blue and/or
red residuals are non-Gaussian in most cases.  4 out of 25 sources
detected in the C$^{18}$O(2$-$1) line show weak (3-4$\sigma$) wing
emission of width 4.5$-$8\,km\,s$^{-1}$ (Table 3) which mostly is seen
from the red or blue sides of the profiles.  In the
H$^{13}$CO$^+$(1$-$0) line the wing emission is seen in 2 out of 17
objects detected (Table 3).  25.410+0.105 is a peculiar source showing
broad (7$-$8\,km\,s$^{-1}$) and symmetric wings in both lines
(Fig. A.1).

In contrast, the optically thick lines show more frequent
absorption dips, multiple components and wings. In several cases
identification of the wings is difficult. The $^{13}$CO(2$-$1)
profiles are especially complex; commonly they are fit by 2-5
Gaussians. These profiles can be interpreted as multiple emitting
regions along the same line of sight. The $^{13}$CO(2$-$1) lines 
show evidence of wings in only 3 objects (Table 3).

%====================================================================================
\begin{table}
\caption{Statistics of wing occurrence.  Entries marked Y or N indicate 
  symmetric wings and no wings respectively, Yb or Yr indicate wing
  emission seen from the blue and red sides of the profiles,
  respectively. An interrogation  point (?) indicates a tentative wing
  and the absence of entry indicates no observation.}
\begin{tabular}{l c c c c c c }
\hline
Source &C$^{18}$O&H$^{13}$CO$^+$&\hskip-2mm$^{13}$CO&\hskip-1mmHCO$^+$&\hskip-1mmCS   &\hskip-1mmCS   \\
       &  (2-1)  & (1-0)        &\hskip-2mm  (2-1)  &\hskip-1mm (1-0) &\hskip-1mm(2-1)&\hskip-1mm(3-2)\\
\hline
21.407$-$0.254 & N &   &\hskip-0.5mm N &   & N &    \\
22.335$-$0.155 & N & N &\hskip-0.5mm N & Y & N & N  \\
22.357$+$0.066 & Yb& N &\hskip-0.5mm N & N & N & N  \\
23.707$-$0.198 & N & N &\hskip-0.5mm N & N & N & N  \\
23.966$-$0.109 & N & Yr&\hskip-0.5mm N & Y & Y & Y  \\
24.147$-$0.009 & N & N &\hskip-0.5mm N & Y & N & N  \\
24.541$+$0.312 & Y & N &\hskip-0.5mm N & Y & N & N  \\
24.635$-$0.323 & Yr& N &\hskip-0.5mm Y & Y & Y & Y  \\
25.410$+$0.105 & N & Y &\hskip-0.5mm Y & Y & Y & Y  \\
26.598$-$0.024 & N & N &\hskip-0.5mm N & N & N & N  \\
27.221$+$0.136 & N &   &\hskip-0.5mm N & N & N &    \\
28.817$+$0.365 & N &   &\hskip-0.5mm N & Y & Y &    \\
30.316$+$0.069 & N & N &\hskip-0.5mm N & Y & N & N  \\
30.398$-$0.297 & Yb& N &\hskip-0.5mm ? & Y & Y & N  \\
31.056$+$0.361 & N &   &\hskip-0.5mm N & N & N &    \\
31.156$+$0.045 & N & N &\hskip-0.5mm ? & Y & Y & Y  \\
31.585$+$0.080 & N &   &\hskip-0.5mm N & N & N &    \\
32.966$+$0.041 & N & N &\hskip-0.5mm ? & N & N & N  \\
33.648$-$0.224 & N &   &\hskip-0.5mm N & N & N & N  \\
33.980$-$0.019 & N & N &\hskip-0.5mm N & Y & Y & Y  \\
34.753$-$0.092 & N &   &\hskip-0.5mm ? & Y & N &    \\
35.791$-$0.175 & N & N &\hskip-0.5mm ? & Y & Y & Y  \\
36.115$+$0.552 & N & N &\hskip-0.5mm N & Y & N & N  \\
36.704$+$0.096 & N &   &\hskip-0.5mm N & Y & N & N  \\
37.030$-$0.039 &   &   &\hskip-0.5mm N & N &   &    \\
37.479$-$0.105 &   &   &\hskip-0.5mm ? & N &   &    \\
37.600$+$0.426 &   &   &\hskip-0.5mm Y & Y &   &    \\
39.100$+$0.491 & N & N &\hskip-0.5mm N & Y & Y & Y  \\
\hline 
\end{tabular}
\label{wings}
\end{table}
%====================================================================================  

The HCO$^+$(1$-$0) lines are also complex, often exhibiting two or more
components or broad line wings (Fig. A.1). They appear to consist
of the superposition of several emitters seen along the line of sight
or of (self)absorption by cooler gas on the near side of the
source. Wings are identified in 17 out of 27 detections (Table
4). The wing full width ranges from 6 to 20\,km\,s$^{-1}$ with a mean
value of 10.3$\pm$3.3\,km\,s$^{-1}$.

Evidence for wings is seen in the CS(2$-$1) transition for
9 out of 25 sources and in the CS(3$-$2) transition for 7 out of 
19 sources (Table 3). Their full widths are from 8 to 19\,km\,s$^{-1}$.

We conclude that 64\% (18/28) of the sources
show residual line wings at least in one line when a Gaussian profile
is used to fit the CO, HCO$^+$ and CS molecular lines.
Detection of the wings may indicate molecular outflows from the MYSOs
identified by methanol masers but we caution that such detections
based on our data alone are only tentative.

\begin{table*}
\caption {Derived properties}
\begin{tabular}{l l l l c c c c c c c }
\hline
       &                &               &               &
                    &                & \multicolumn{2}{c} {30K} &&\multicolumn{2}{c} {60K}  \\
\cline{7-8} \cline{10-11}     
Source &  $V_{\rm sys}$ & $d_{\rm near}$& $d_{\rm far}$ &
$N$(H$^{13}$CO$^+$) & $N$(C$^{18}$O) & log$n_{\rm H_2}$ & log$N$(CS) &&  log$n_{\rm H_2}$ & log$N$(CS) \\
       & (km\,s$^{-1}$) &  (kpc)     &    (kpc)    & 
       (10$^{12}$cm$^{-2}$) & (10$^{15}$cm$^{-2}$)
       &(cm$^{-3}$)   &  (cm$^{-2}$) &&(cm$^{-3}$)   &  (cm$^{-2}$)    \\
\hline
21.407$-$0.254 & 90.7  & 6.0  & 10.4 &-    &  3.8  &-  &-  &&- &- \\
22.335$-$0.155 & 30.9  & 2.4  & 14.7 & 2.1 &  3.7  &6.15$\pm$0.15 &  14.52$\pm$0.10 && 5.91$\pm$0.12 & 14.68$\pm$0.06 \\
22.357$+$0.066 & 84.2  & 5.2  & 10.6 & 2.2 & 19.1  &5.48$\pm$0.09 &  14.70$\pm$0.11 &&5.27$\pm$0.13 & 14.52$\pm$0.21 \\
23.707$-$0.198 & 68.9  & 5.1  & 10.5 & 3.2 & 13.2  &5.93$\pm$0.08 &  13.74$\pm$0.14 &&5.57$\pm$0.08 & 13.49$\pm$0.07 \\
23.966$-$0.109 & 72.7  & 4.2  & 11.6 & 5.1 &  9.0  &$>$6.7  &  14.73$\pm$0.57 && $>$6.5  & 15.16$\pm$0.52 \\
24.147$-$0.009 & 23.1  & 2.0  & 14.5 & 1.4 &  2.1  &5.61$\pm$0.08  & 14.51$\pm$0.18 && 5.42$\pm$0.12 & 14.60$\pm$0.56 \\ 
24.541$+$0.312 & 107.8 & 7.0  &  9.5 & 1.5 &  4.6  &-  &- &&- &- \\
24.635$-$0.323 & 42.7  & 3.7  & 13.1 & 4.7 &  7.6  &$>$6.7   & 14.61$\pm$0.38 && 6.39$\pm$0.12  & 14.53$\pm$0.23 \\ 
25.410$+$0.105 & 96.0  & -    &  9.5 & 3.4 &  7.0  &6.42$\pm$0.11 & 14.40$\pm$0.08 && 6.22$\pm$0.07  & 14.53$\pm$0.20 \\
26.598$-$0.024 & 23.3  & 1.8  & 13.4 & 1.8 & 32.6  &$>$6.9   & 14.54$\pm$0.42 && $>$6.5   & 14.86$\pm$0.28 \\
27.221$+$0.136 & 112.6 & -    &  8.0 &-    &  9.4  &-  &- &&- &- \\
28.817$+$0.365 & 87.0  & 5.5  &  9.4 &-    &  5.3  &-  &- &&- &- \\
30.316$+$0.069 & 45.3  & 2.8  & 12.2 & 2.2 &  3.3  &$>$6.9   & 14.77$\pm$0.18&& 6.28$\pm$0.19  & 14.59$\pm$0.09 \\
30.398$-$0.297 & 102.4 & 6.0  &  8.5 & 1.6 &  3.7  &6.12$\pm$0.10  &  14.83$\pm$0.08 &&- &- \\
31.056$+$0.361 & 77.6  & -    &  9.6 &-    &  2.9  &-  &- &&- &- \\
31.156$+$0.045 & 38.9  & 2.7  & 11.9 & 2.2 &  4.8  &6.06$\pm$0.04  &
       14.11$\pm$0.06 && 5.74$\pm$0.06  & 14.64$\pm$0.13 \\
31.585$+$0.080 & 96.0  & 5.4  &  8.1 &-    & 11.8  &-  &- &&- &- \\
32.966$+$0.041 & 83.4  & 5.4  &  8.9 & 1.3 &  4.2  &-  &- &&4.39$\pm$0.11  & 15.73$\pm$0.16 \\
33.648$-$0.224 & 61.5  & -    & 10.4 &-    &  2.1  &-  &- &&- &- \\
33.980$-$0.019 & 61.1  & 3.5  & 10.6 & 2.5 &  4.7  &-  &- &&4.52$\pm$0.21  & 15.76$\pm$0.13  \\
34.753$-$0.092 & 51.1  & 3.1  & 11.0 &-    &  1.4  &-  &- &&- &- \\
35.791$-$0.175 & 61.9  & 4.6  & 10.3 & 2.4 &  3.0  &-  &- &&- &- \\
36.115$+$0.552 & 76.0  & 4.9  &  9.0 & 1.9 &  8.1  &-  &- &&$>$6.8$^{90}$& 15.60$\pm$0.19$^{90}$ \\
36.704$+$0.096 & 59.8  & 4.6  & 10.4 &-    &  0.9  &-  &- &&- &- \\
37.030$-$0.039 & 80.1  & 5.0  &  8.3 &-    &-      &-  &- &&- &- \\
37.479$-$0.105 & 59.1  & -    &  9.5 &-    &-      &-  &- &&- &- \\
37.600$+$0.426 & 90.0  & 6.5  &  7.5 &-    &-      &-  &- &&- &- \\
39.100$+$0.491 & 23.1  & 1.0  & 14.7 & 2.0 &  2.9  &-  &- && 6.58$\pm$0.08$^{90}$ & 14.81$\pm$0.08$^{90}$ \\
\hline
\end{tabular}
\label{parameters}

$^{90}$ values for kinetic temperature 90\,K
\end{table*}

\section{Derivation of physical parameters}

\subsection{Column densities}
In order to estimate the column density of H$^{13}$CO$^+$
from the observed HCO$^+$(1$-$0) and H$^{13}$CO$^+$(1$-$0)
line parameters, we follow the procedure  outlined in 
Purcell et al. (\cite{purcell06}) and references therein.
Briefly, the main assumptions made are: 
(i) HCO$^+$(1$-$0) is optically thick and H$^{13}$CO$^+$(1$-$0) 
is optically thin.
(ii) Both lines form in the same gas and share the same
excitation temperature.
(iii) The excitation temperature is equal to the rotational
temperature.
(iv) The gas is in local thermodynamic equilibrium.
(v) The beam filling factor is one for both lines.

The derived H$^{13}$CO$^+$ column density, $N$(H$^{13}$CO$^+$), 
(Table 4) ranges from $1.3 - 5.1\times10^{12}$\,cm$^{-2}$ 
and the median value is $2.2\times10^{12}$\,cm$^{-2}$.
We derive a value of $N$(H$^{13}$CO$^+$) a factor of 4 smaller
than the value found by Purcell et al. (\cite{purcell06}) for two of
the sources common to both samples, 22.357$+$0.066 and 23.707$-$0.198.  
This is probably because Purcell et al. applied corrections for 
self-absorption, leading to higher estimates of the  HCO$^+$(1$-$0)
line intensities and lower optical depths, compared with our study.
We adopt an abundance ratio of [H$^{13}$CO$^+$/H$_2$]=3$\times$10$^{-11}$ 
(Girart et al.\,\cite{girart00}), from which we obtain the
H$_2$ column density from $4.3 - 17.0\times10^{22}$\,cm$^{-2}$
with the median value of $7.3\times10^{22}$\,cm$^{-2}$.

We apply the same method to estimate the column density
of C$^{18}$O, $N$(C$^{18}$O), from the line parameters of 
$^{13}$CO(2$-$1) and C$^{18}$O(2$-$1), assuming that 
$^{13}$CO(2$-$1) is optically thick and C$^{18}$O(2$-$1) 
is optically thin. For our sample $N$(C$^{18}$O) is
0.9$-$32.6$\times$10$^{15}$\,cm$^{-2}$ (Table 4) with 
the median value of 4.6$\times10^{15}$\,cm$^{-2}$. 
The temperature varies between 10 and 30\,K. 
The resulting H$_2$ column density ranges from 
$5.4\times10^{21} - 1.9\times10^{23}$\,cm$^{-2}$ 
for an abundance ratio [C$^{18}$O]/[H$_2$]=1.7$\times$10$^{-7}$
(Frerking et al.\,\cite{frerking82}).

We conclude that the CO and HCO$^+$ data provide consistent
estimates of the column density of H$_2$ towards 
the methanol maser sources. The range of $N$(H$_2$)
derived here is in good agreement with that reported
for high-mass protostar candidates associated with methanol
masers; $3\times10^{22} - 2\times10^{23}$\,cm$^{-2}$
(Codella et al.\,\cite{codella04}; Minier et al.\,\cite{minier05}; 
Purcell et al.\,\cite{purcell06}). However, it is significantly lower than 
$N$(H$_2$)$\ge 4\times10^{23}$\,cm$^{-2}$ reported in
some earlier works (e.g. Churchwell et al.\,\cite{churchwell92}) 
for ultra-compact HII regions. This discrepancy is likely due to
the temperatures of 10$-$30\,K derived here which is significantly
lower than $\ge 100$\,K assumed in Churchwell et al. (\cite{churchwell92}).

We notice that a dispersion of the $N$(C$^{18}$O)
is a factor of 7 larger than that of the $N$(H$^{13}$CO$^+$) (Table 4).  
In two sources 22.357$+$0.066 and 26.598$-$0.024 the $N$(C$^{18}$O) is
extremely large ($>1.9\times 10^{16}$\,cm$^{-2}$). In consequence, the 
values of $N$(H$_2$) derived from the C$^{18}$O is a factor of 1.5
and 3.1, respectively, higher that those derived from the H$^{13}$CO$^+$.
This discrepancy suggests that the methanol masers in these sources
probe regions with the abundance ratio of $^{13}$CO/C$^{18}$O
significantly lower than a typical ratio of 6.5$-$7 
(Frerking et al.\,\cite{frerking82}; Beuther et al.\,\cite{beuther00}).
A decrease of $^{13}$CO/C$^{18}$O ratio is predicted in the PDR model
in a clumpy cloud; in small clumps the C$^{18}$O molecule is nearly 
completely photodissociated whilst it is protected from
photodissociation in large clumps (Beuther et al.\,\cite{beuther00}
and references therein). 
Object 26.598$-$0.024 with the highest value of $N$(C$^{18}$O) is also
a candidate infall object (Sect. 4.2.1) and one can speculate that it is
the youngest methanol maser in our sample; the maser emission forms
in large clumps at nearly systemic velocity. 
Another explanation of low $^{13}$CO/C$^{18}$O intensity 
ratio can be that our 11\arcsec ~beam probes the methanol maser sites 
where the C$^{18}$O cores did not coincide with the$^{13}$CO cores.
This observational fact is well documented in Brand et
al.\,(\cite{brand01}) at least for their sources Mol 98 and Mol 136
(see their Fig. 5). Furthermore, the C$^{18}$O emission is less
extended than the $^{13}$CO emission; by a factor of $\sim 3-5$ for 
common source 35.791$-$0.175. This explanation seems to be less
plausible as a similar effect can be observed for HCO$^+$ and 
H$^{13}$CO$^+$ lines. 

\subsection{Gas density and temperature}
We used the escape-probability modelling code RADEX
on-line{\footnote{http://www.strw.leidenuniv.nl/moldata/radex.php}}
to estimate the density and temperature of the gas required for the
observed line temperature ratios of CS and C$^{34}$S. Because these
parameters cannot be derived independently for diatomic molecules
(Schilke et al.\,\cite{schilke01}) we calculate the models for 30,
60 and 90\,K with gas number densities of $10^4-10^7$\,cm$^{-3}$, 
CS column densities of $10^{12}-10^{17}$\,cm$^{-2}$ and linewidth 
of 1\,km\,s$^{-1}$.  We performed the calculations for the 16 sources 
for which all three CS lines were detected and we assumed that beam 
dilution is comparable for all these transitions. We used a $\chi^2$
minimization procedure to fit the models to the observed line
ratios. The derived parameters are listed in Table 4. 
We found equally reasonable fits for 10 sources using models at kinetic
temperatures of both 30 and 60\,K.  Five sources have good fits only
for a single kinetic temperature.  We could not find a satisfactory 
fit for the source 35.791$-$0.175 as its CS(2$-$1) and CS(3$-$2) lines 
are strongly self-absorbed (Fig. A.1) and thus its
line ratios are poorly constrained.

Using a temperature of 60\,K the average logarithmic number
density is 5.7$\pm$0.7 and the average logarithmic column density of
CS is 14.7$\pm$0.6 for the sample. These values are consistent with
5.9 and 14.4, respectively, reported for a large sample of massive
star formation sites selected by the presence of H$_2$O masers 
(Plume et al.\,\cite{plume97}). Our estimates are also in good 
agreement with those based on the nine-point CS maps of high-mass 
protostellar candidates (Beuther et al.\,\cite{beuther02a}; 
Ossenkopf et al.\,\cite{ossenkopf01}) and calculated with more 
sophisticated models. Taking the CS fractional abundance as
$\sim 8\times 10^{-9}$ (Beuther et al.\,\cite{beuther02a}) 
our estimate of the CS column density implies a mean $N$(H$_2$) of
$6.3\times10^{22}$\,cm$^{-2}$ which is in very good agreement
with the estimates based on CO and HCO$^+$ data (Sect. 5.1).

Our C$^{34}$S data are less useful to estimate the gas density and
temperature because the line ratios are poorly constrained for most of
the targets.  26.598$-$0.024 is the only source for which we are
able to determine C$^{34}$S line ratios but the results are
inconsistent with those obtained from the CS data.   This indicates
that the escape probability model provides only a crude estimate to
the physical parameters and the assumption of homogeneous parameters
across the cloud is not fulfilled (Ossenkopf et al.\,\cite{ossenkopf01}).

\section{Discussion}

%===================================================================================
%===================================================================================
\begin{figure}[t]
    \resizebox{\hsize}{!}{\includegraphics[angle=0]{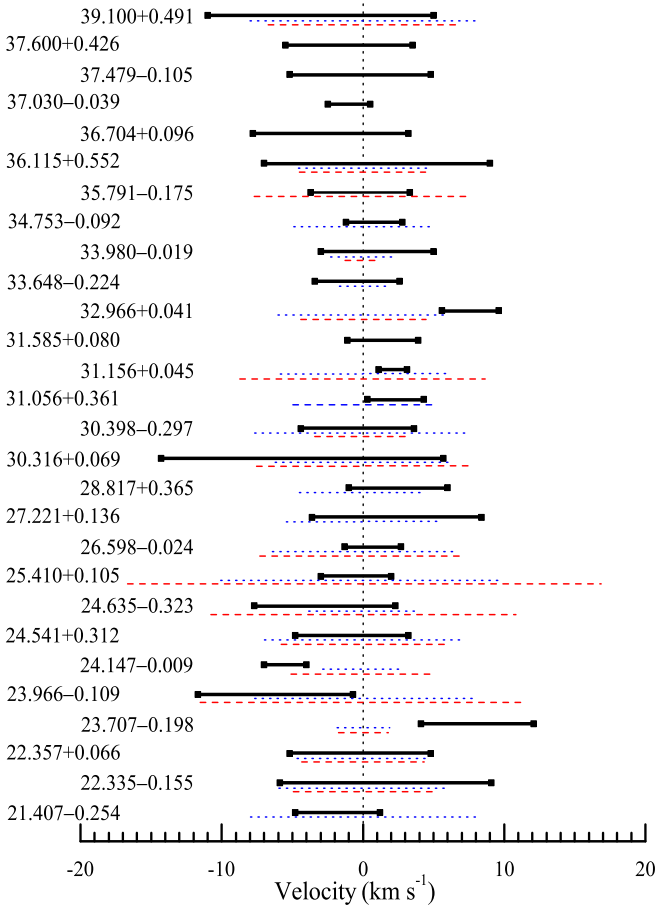}}
   \caption{Comparison between the velocity ranges of 6.7\,GHz
   methanol maser (thick bars)(Szymczak et al.\,\cite{szymczak02}) and
   $^{13}$CO (dotted bars) and HCO$^+$ (dashed bars) line wings.
   The dotted vertical line marks the systemic velocity.}
   \label{vel_comparison}
\end{figure}
%====================================================================================
%====================================================================================

\subsection{Kinematics}
The present survey reveals new information regarding the kinematics
of molecular gas surrounding massive forming stars. In the following
we attempt to answer the question of whether the 6.7\,GHz methanol maser
and the thermal molecular lines arise from similar or different 
kinematic regimes. 

The velocity ranges of 6.7\,GHz methanol masers, $^{13}$CO and
HCO$^+$ line wings are plotted in Fig. 3. This plot clearly
shows that the systemic velocity derived in this study (Table 4)
is in good agreement with the methanol maser central velocities,
$V_{\rm m}$, derived from Szymczak et al.\,(\cite{szymczak02}).
We note that in many sources $V_{\rm m}$ does not coincide with the
peak maser velocity $V_{\rm p}$.
The average value of $V_{\rm m}-V_{\rm sys}$ is 
$0.04\pm$0.60\,km\,s$^{-1}$. The difference is less than 
3\,km\,s$^{-1}$ for 23 sources
(82\%). $V_{\rm m}$ is offset by $>$4 and $\le$8.1\,km\,s$^{-1}$ with
respect to $V_{\rm sys}$ in 5 sources (18\%), 23.707$-$0.198,
23.966$-$0.109, 24.147$-$0.009, 30.316$+$0.069 and 32.966$+$0.041
(Figs. 3 and A.1). This does not necessarily imply that the
different species arise from separate regions along the same line of
sight.  Two of the sources, 24.147$-$0.009 and 32.966$+$0.041, have
ranges of maser emission $\Delta V_{\rm m}\le$4\,km\,s$^{-1}$ which is
a factor of two narrower than the mean value of
8.3$\pm$0.9\,km\,s$^{-1}$ for the sample but this could be simply an
effect of inhomogeneous conditions in molecular clumps; the maser
emission is sustained in one or a few clumps of sizes a
few$\times10^{15}$\,cm (Minier et al.\,\cite{minier00}).   The effect
of clumping is clearly seen even in regular structures (Bartkiewicz
et al.\,\cite{bartkiewicz05}).  The other three sources exhibit 
maser emission at velocities which differ from the systemic velocity
by less than 4\,km\,s$^{-1}$.  In source 30.316+0.069 the maser
spectrum is double (Szymczak et al.\,\cite{szymczak00}) and one of the
peaks near 49\,km\,s$^{-1}$ is close to the systemic velocity of
45.3\,km\,s$^{-1}$, so that the maser emission related to the thermal
molecular lines has a width of about 6\,km\,s$^{-1}$.  We conclude,
$V_{\rm m}$ is a reliable estimator of the systemic velocity, with an
accuracy better than 3\,km\,s$^{-1}$, for most of the sources in our
sample.

The overlap between the velocity ranges of the methanol masers
and the $^{13}$CO/HCO$^+$ line wings is remarkable. 
Figure 4 shows a histogram of the ratio of methanol maser
velocity spread, $\Delta V_{\rm m}$, to HCO$^+$ line wings spread. 
This ratio ranges from 0.2$-$6.7 and the median value is 1.3. 
Similar trends are observed in the ratio of $\Delta V_{\rm m}$ to 
$^{13}$CO line wings spread. In 12 out of 23
sources where we detected $^{13}$CO/HCO$^+$ wings, $\Delta
V_{\rm m}$ falls entirely within the wing velocity ranges and
in 9 sources there is an overshoot of $\le$4\,km\,s$^{-1}$. 
The $^{13}$CO/HCO$^{+}$ line wings appear to provide a good 
indication of the presence of outflow and their  widths can serve 
as an approximate measure of outflow velocities. The present 
observations used beamwidths of 11\arcsec\, and 27\arcsec\, for 
$^{13}$CO and HCO$^+$ lines, respectively, which samples a small 
fraction of the molecular cloud, centred on the methanol maser position.  

The outflow velocity can be reliably estimated from 
these data only for the fortunate case when the axis of outflow lies 
along the line of sight. One source in our sample,  25.410+0.105, 
has been mapped in the $^{12}$CO(2$-$1) line by 
Beuther et al.\,(\cite{beuther02b}) who measured a wing velocity range of 
14\,km\,s$^{-1}$, which is comparable with our estimate. In this
object the maser emission, with velocity width of 5\,km\,s$^{-1}$,
is closely centered on the systemic velocity. The velocity ranges of the
$^{13}$CO and HCO$^+$ wings are 11 and 18\,km\,s$^{-1}$,
respectively.  This indicates that the maser
emission traces a small portion  of the kinematic regime of the 
$^{13}$CO and HCO$^+$ lines or it is completely unrelated.
Fig. 3 indicates that sources 21.407$-$0.254, 26.598$-$0.024,
31.156+0.045 and 35.791$-$0.175 share similar properties with 
25.410+0.105. 
VLBI observations of 35.791$-$0.175 
(Bartkiewicz et al.\,\cite{bartkiewicz04}) support the above 
interpretation. In this object the 6.7\,GHz methanol maser emission 
appears to come from part of a circumstellar disc.

Our sample contains 4 objects (23.707$-$0.198, 24.147$-$0.009,
32.966+0.041, 36.115+0.552) for which the velocity range of the
maser emission is very similar to or slightly overshoots that of
the $^{13}$CO/HCO$^{+}$ line wings. If we assume that the
width of $^{13}$CO and HCO$^{+}$ line wings is a measure of the
outflow velocity, in these objects the 6.7\,GHz methanol masers
arise in outflows.  This scenario appears to be supported by VLBI
observations of 36.115+0.552 (Bartkiewicz et
al.\,\cite{bartkiewicz04}); the maser emission comes from two well
separated regions which probably represent a bipolar outflow.  
In this case the methanol maser traces the same or a very similar 
kinematic regime as that of the $^{13}$CO and HCO$^+$ lines.

Sources 22.355$-$0.155 and 27.221+0.136 appear to posses complex
kinematics in the regions where the methanol masers operate. 
A close inspection of their 6.7\,GHz spectra 
(Szymczak et al.\,\cite{szymczak02}) suggests
that some spectral features arise from the inner parts of the
molecular cloud whilst other features form in outflows.
VLBI studies of maser emission and detailed measurements of the 
kinematic properties of the molecular emission are needed 
to verify this suggestion.

%===================================================================================
%===================================================================================
\begin{figure}[t]
    \resizebox{\hsize}{!}{\includegraphics[angle=0]{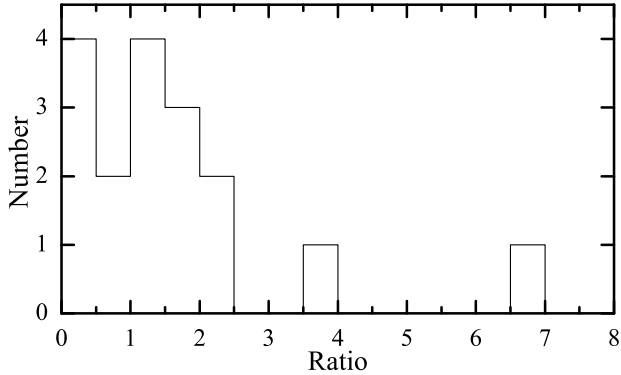}}
   \caption{Histogram of the ratio of methanol maser velocity spread
   to HCO$^+$ line wings spread.}
   \label{vel_comparison}
\end{figure}
%====================================================================================
%====================================================================================

\subsection{Implications for the evolutionary status }
One of the important findings of our observations is the
detection of considerable number of sources with line wings.
We identified residual line wings in 18 out of 28 sources when
a Gaussian profile was used to fit the CO, HCO$^+$ and CS molecular 
lines. The line wings appear to be the best indicators of outflow 
motions in most cases. The presence of line wings in about 64\% of
sources in the sample suggests a close association of the methanol 
masers with the evolutionary phase when outflows occur.
This result is consistent with that reported by Zhang et
al. (\cite{zhang05}). They mapped the CO(2$-$1) line in a sample
of 69 luminous IRAS point sources and found that  
about 60\% of them were associated with outflows.
However, with the present data we cannot resolve whether 
the methanol maser sites and the outflows have a common origin.
Because of clustering in high mass star formation 
(e.g. Beuther et al.\,\cite{beuther02a}) it is 
possible that some masers in the sample are not actually 
associated with outflowing sources.

Codella et al. (\cite{codella04}) proposed an evolutionary
sequence for UCHII regions in which the earliest phase 
is marked by maser emission and molecular outflows not yet large 
enough to be detected with single-dish observations. 
The present data suggest that our sources are slightly more
evolved because several of them show evidence of outflows.
Their age therefore seems to be less than a few 
10$^4$\,yr (Codella et al.\,\cite{codella04}) which is
consistent with a statistical estimate of $3-5\times10^4$\,yr
for the methanol maser lifetime (van der Walt\,\cite{vanderwalt05}).

\subsection{Constraints on  maser models}
The present study allows us to refine the range of physical
conditions required to produce strong methanol masers at
6.7\,GHz. Theoretical modelling by Cragg et al. (\cite{cragg02})
demonstrated that a maser line of 1\,km\,s$^{-1}$ width attains
a peak brightness temperature of $\sim$10$^{11}$\,K for a dust
temperature $>$100\,K and a methanol column density
$>5\times10^{15}$\,cm$^{-2}$. They found that methanol masers can
be produced under a wide range of the physical conditions.  
In fact, for a methanol fractional abundance from $3\times10^{-8}$ to
$10^{-5}$, masing is predicted for the gas density range $10^5 -
2\times 10^8$\,cm$^{-3}$ and the methanol column density range
$5\times10^{15} - 2\times10^{18}$\,cm$^{-2}$ (Cragg et
al.\,\cite{cragg02}).  The gas density inferred from our observations
is between 10$^5$ and 10$^7$\,cm$^{-3}$; higher values
($>10^7$\,cm$^{-3}$) are less probable.  The hydrogen column density
from $10^{22}$ to $2\times10^{23}$\,cm$^{-2}$, inferred here, 
translates well into the above range of methanol column densities for
methanol fractional abundances of $5\times10^{-7} - 10^{-5}$.  This
suggests that 6.7\,GHz maser emission is less probable in 
environments with a lower methanol fractional abundance of the order
of $10^{-8}$.  We conclude that our study well refines a range of the
input parameters of Cragg et al.'s maser model.  Specifically, a high
methanol fractional abundance of $>5\times10^{-7}$ is required
whilst a gas density $<10^7$\,cm$^{-3}$ is sufficient for the
production of methanol masers.

\section{Conclusions}
We have observed 10 transitions of HCO$^+$, CO and CS isotopomers at
millimetre wavelengths in order to characterize the physical
conditions in a sample of 28 MYSOs identified by the presence 
of methanol masers. No other preconditions were involved in the sample
selection. The observations were centred at maser positions known 
with a sub-arcsecond accuracy. 
The main conclusions of the paper are summarized as follows:

(1) The systemic velocity determined from the
optically thin lines C$^{18}$O(2$-$1), H$^{13}$CO$^+$(1$-$0),
C$^{34}$(2$-$1) and C$^{34}$(3$-$2) agrees within $\pm$3\,km\,s$^{-1}$
with the central velocity of the methanol maser emission 
for almost all the sources.

(2) 18 out of 28 sources show residual line wings at least in one line
when a Gaussian function was used to fit the CO, HCO$^+$ and CS
lines. 
Detection of the line wing emission suggests the presence of 
molecular outflows in these sources.
Their occurrence needs to be confirmed by mapping observations.

(3) Comparison between the kinematics of the methanol masers and of 
the thermal molecular lines reveals that they trace a wide range of 
molecular cloud conditions. 
In some objects the maser emission occurs in a narrow velocity
range centered at the systemic velocity, which may indicate that the
innermost parts of a molecular cloud or a circumstellar disc is the
site of maser emission.  In other objects the velocities of maser
features are very similar to, or slightly overshoot, the velocity ranges
of the thermal molecular line wings, suggesting that the masers
arise in outflows. There are also objects where the maser
emission reveals more complex kinematics.

(4) The column density of H$_2$ derived from the CO and HCO$^+$ lines
are between $10^{22}$ and $2\times 10^{23}$\,cm$^{-2}$.  
We use our measurements of the intensity ratios of the CS lines 
to infer that methanol masers arise from regions with a gas density 
of $10^5 - 10^7$\,cm$^{-3}$, a kinetic temperature of $30 - 100$\,K 
and a methanol fractional abundance of $5\times10^{-7} - 10^{-5}$. 
This represents a significant refinement to the input parameters of
methanol maser models.

\begin{acknowledgements}
 We like to thank the staff of the IRAM 30\,m telescope
 for help with the observations and the unknown referee for
 helpful comments. 
 This work has been supported by the Polish MNiI grant 1P03D02729. 
\end{acknowledgements}

\end{document}